# Cosmic Magnetic Fields

Elisabete M. de Gouveia Dal Pino

*Instituto de Astronomia, Geofísica e Ciências Atmosféricas, Universidade de São Paulo,*
*Rua do Matão 1226, Cidade Universitária, CEP 05508-900, São Paulo, SP, Brazil*
*e-mail: dalpino@astro.iag.usp.br*

**Abstract.** Most of the visible matter in the Universe is in a plasma state, or more specifically is composed of ionized or partially ionized gas permeated by magnetic fields. Thanks to recent advances on the theory and detection of cosmic magnetic fields there has been a worldwide growing interest in the study of their role on the formation of astrophysical sources and the structuring of the Universe. In this lecture, I will briefly review the importance of the cosmic magnetic fields both from a theoretical and from an observational perspective, particularly focusing on stellar and compact objects, the interstellar medium and star formation regions, and on galaxies, clusters of galaxies, and the primordial Universe.

## INTRODUCTION

Few decades ago cosmic magnetic fields were generally regarded as unimportant. Only a few scientists, like Alfvén, Biermann, Chandrasekhar, and Parker, realized the potential role of the magnetism in the Universe. Since then, our view has changed considerably. We know now that most of the visible matter in the Universe is in a plasma state, or more specifically is composed of ionized or partially ionized gas permeated by magnetic fields. There seems to be now no doubts that magnetic fields play a crucial role in: star formation, solar and stellar activity, pulsars, accretion disks, formation and stability of jets, formation and propagation of cosmic rays, and stability of galactic disks. They are also probably crucial in: the interstellar medium (ISM) dynamical evolution, in molecular clouds, supernova remnants, proto-planetary disks, and planetary nebulae, but its importance is still not well understood in: stellar evolution, halos of galaxies, galaxy evolution, and structure formation in the early Universe (e.g., [1]). In this lecture, I will *very briefly* review the importance of the cosmic magnetic fields, particularly focusing on compact objects, the ISM and star formation regions, and on galaxies, clusters, and the early Universe. I will also briefly discuss the potential origin of the cosmic magnetic fields.

Before starting this review, let me consider the magnetic induction equation that describes the magnetic field evolution in a fluid:

$$\frac{\partial \boldsymbol{B}}{\partial t} = \boldsymbol{\nabla} \times (\boldsymbol{v} \times \boldsymbol{B}) + \frac{1}{4\pi\sigma}\nabla^2 \boldsymbol{B} \qquad (1)$$

Where $\sigma$ is the electric conductivity. The first term of the RHS accounts for the flux freezing between the ionized gas and the magnetic field lines and the second one accounts for the magnetic diffusion. The comparison of the second term with the one of the LHS gives the diffusion time scale for the magnetic field configuration decay, $t_{diff}(L) = 4\pi\sigma L^2$, where $L$ is the characteristic length scale of the spatial variation of $\boldsymbol{B}$. In most astrophysical systems and environments, the gas is a good conductor and since the typical length scales $L$ are very large, one finds that $t_{diff}(L)$ is generally much longer than the typical dynamical time scales of the systems. This implies that the first term on the RHS (the frozen-in term) generally dominates and the magnetic flux is conserved in most astrophysical plasmas, with the possible exception of magnetic boundary layers or surface discontinuities, like magnetic reconnection regions, where the condition above may be violated. Therefore, as a *golden rule* in Astrophysics, one may say that once magnetic fields are created in cosmic environments it is very difficult to get rid of them!

# HOW DO WE MEASURE COSMIC MAGNETIC FIELDS?

Cosmic magnetic fields are observable through optical polarization of (unpolarized) star light by interstellar dust grains that are aligned with the embedded magnetic field in the ISM; through the Zeeman splitting of spectral emission lines from sources by the ambient field; through linearly polarized radio synchrotron emission produced by relativistic particles accelerated in the magnetic fields of some astrophysical sources like pulsars, supernova remnants or radio galaxies; and through Faraday rotation of the angle between the polarization vector and the line of sight by the ambient magnetic field where the polarized light travels. For more detailed reviews on these observational methods and techniques I refer to Beck 2005a, Cructcher 2005, Magalhães 2005, and references therein.

# MAGNETIC FIELDS IN STARS AND COMPACT OBJECTS

In the case of the Sun, our nearest and, therefore, most investigated astrophysical source, it is well known that its surface phenomena, such as flares and coronal mass ejections (CMEs), are all the result of intense magnetic activity. Magnetic arcs that extend up to $10^4$ km above the surface rise by buoyancy due to convective motions. The sunspots that are located at the feet of these magnetic arcs have magnetic field intensities that can be as large as 2000 G. Solar flares, which correspond to sudden release of energy of the order of $10^{30}$-$10^{32}$ erg in a period of seconds to hours, are caused by violent magnetic reconnection of the lines in the magnetic loops (e.g., Carmichael; Shibata 2005; Silva 2005) and the released energy is responsible for the solar coronal gas heating (up to temperatures of the order of $2 \times 10^6$ K), particle acceleration, and the production of the CMEs.

The magnetic fields are created by dynamo action in the convection layer under the solar atmosphere, and rise up to the surface by magnetic buoyancy effects. In the dynamo process, it is the conversion of the kinetic energy of the convective and turbulent motions (through the so called *alpha* effect) and the conversion of the kinetic energy of the differential rotation motions (through the *omega* effect) into magnetic energy that is responsible for the amplification and continuous regeneration of the poloidal and toroidal components of the solar dipole magnetic field (e.g., Grasso & Rubinstein 2001; Shibata 2005; Guerrero & de Gouveia Dal Pino 2006, and references therein). The same physical processes are expected to occur in several other classes of stars (especially in the cool ones) and similar, but more violent magnetic activity may be occurring also in accretion disks and galactic disks (see below).

## Jets and Accretion Disks

Narrow conical (or cylindrical) supersonic jets that channel mass, momentum, energy and magnetic flux from their parent sources to the outer ambient medium are observed in a wide variety of astrophysical objects, from young stars (YSOs) and compact evolved objects (like galactic black holes or microquasars, and X-ray binary stars), to the nuclei of active radio galaxies (AGNs). Despite their different physical scales (in size, velocity, source mass, and amount of energy transported), they have strong morphological similarities that suggest a universal mechanism for their origin. The currently most accepted model for jet production is based on the magneto-centrifugal acceleration out off a magnetized disk that surrounds the central source and accretes matter into it (Blandford & Payne 1982, Konigl & Pudritz 2000; see also de Gouveia Dal Pino 2005a, b; Raga; and Cerqueira et al. in these Proceedings, for reviews on this model and the jet phenomena). There has been recently some observational and numerical evidences in favor of this model (e.g., Bacciotti et al., Shibata 2005, de Gouveia Dal Pino 2005, 2006), but the origin of the large scale magnetic field which is required for this mechanism to operate at the accretion disk is still uncertain. It could be either *advected* from the ambient medium by the infalling material or could be locally produced by dynamo action in the disk in a similar way to the one previously discussed for the sun and other stars. In this regard, we have recently proposed that the observed X-ray and radio flares events accompanied by ejections of relativistic particles in relativistic jets (e.g., GRS 1915+105) could be produced by violent magnetic reconnection episodes just like on the solar surface. The process would take place when a large scale magnetic field is established by turbulent dynamo in the accretion disk (de Gouveia Dal Pino & Lazarian 2005; de Gouveia Dal Pino 2006).

# MAGNETIC FIELDS IN THE ISM AND STAR FORMATION REGIONS

Observations of the polarized emission of the diffuse ISM of our Galaxy trace structures of pc and sub-pc sizes that carry valuable information about its turbulent nature. There is a straight correlation between turbulence, the average magnetic field and cosmic rays distribution, with their energy densities nearly in equipartition. The MHD turbulence distributes the energy from supernovae explosions, jets and winds from stars in the ISM of the Galaxy and the magnetic fields control the density and distribution of cosmic rays in the ISM and halo of the Galaxy.

While the magnetic fields in the diffuse ISM and in neutral hydrogen clouds (HI clouds) are predominantly turbulent, they tend to be much more regular within the denser molecular clouds where stars form. Polarization maps of the cores of these clouds are consistent with strong regular fields, ~ 100 µG to few mG (Crutcher 2005). The observations also indicate thermal to magnetic pressure ratios $P_{th}/P_B \approx 0.29$ and 0.04, and turbulent to magnetic pressure ratios $P_t/P_B \approx 1.3$ and 0.7, in the diffuse ISM and the molecular clouds cores, respectively. These figures indicate a predominance of compressible turbulence on the dynamics of the diffuse ISM and of the magnetic fields on the molecular clouds, and suggest the following potential scenario for structure and star formation in the ISM: molecular clouds are possibly formed by the aggregation of HI clouds. Thus, compressible turbulence must dominate the aggregation process over both gravity and magnetic fields, sometimes forming self-gravitating clouds. Since the magnetic fields are strong enough even in the diffuse ISM medium, the aggregation must be primarily along magnetic flux tubes as magnetic pressure resists to compression in the direction perpendicular to the field.

Residual effects of turbulence are seen in self-gravitating molecular clouds, but they appear to be magnetically supported, with the condition $\rho g \approx \nabla P_B$ approximately satisfied. When a cloud core becomes gravitationally unstable, ambipolar diffusion of the neutral component of the gas through the magnetic field lines must drive its gravitational collapse and star formation in a fast (~few free-fall times) timescale. The magnetic field will be also essential for removing the angular momentum excess from the protostellar cloud during this process, causing the so called magnetic braking that will conduct the protostar to its final collapse into a stable star (Crutcher 2005; Li & Shu 1996).

## MAGNETIC FIELDS IN THE MILKY WAY AND OTHER GALAXIES

As generally observed in other galaxies with spiral morphology, the large scale magnetic field in our galaxy, the Milky Way, seems to approximately follow the density spiral arms. However, Faraday rotation measurements (RM) from pulsars radio emission suggest the existence of multiple reversals along the galactic radius that have not been detected so far in any external galaxy. To account for large-scale reversals, a *bisymmetric* magnetic spiral would be required (a configuration that is not predicted by a dynamo process), but the *RM* data from pulsars are still scarce. Furthermore, some of these field reversals may not be of galactic extent, but due to local field distortions or magnetic loops of the anisotropic turbulent field component. In summary, the pattern of the large-scale regular magnetic field of the Milky Way is still unknown (Beck 2005b). A combination of cosmic-ray energy density measurements with radio synchrotron data yields a local strength of the total field of the order of 6 µG and 10 µG in the inner Galaxy, which are similar to the values in other galaxies (see below).

Large-scale spiral patterns of regular field are observed in grand-design, flocculent, and even in some irregular galaxies. In grand-design galaxies the regular fields are aligned parallel to the optical spiral arms, with the strongest regular fields (highest polarization) in inter-arm regions, sometimes forming *magnetic spiral arms* between the optical arms. This is an indication that the magnetic fields do not require the assistance of the density spiral arms to develop.

Regular fields are strongest in inter-arm regions ($B \approx 15$ µG), while turbulent fields are strongest within the spiral arms ($B \approx 20$ µG) due to the presence of intense star formation, stellar winds, jets, and SN shocks there. Processes related to star formation tangle the field lines, so that smaller polarization is observed in star-forming regions.

Most of the field structures in the galaxies require a superposition of several dynamo modes to explain their origin and amplification. The turbulent dynamo (Subramanian 1998) considers a standard dynamo model acting in different scales of multiple cells and these micro-dynamos combine to build a large scale magnetic field. In some cases (e.g., M31), Faraday rotation of the polarized radiation reveals patterns which are signatures of large-scale regular, coherent fields in the galactic disks that could not be generated by direct compression or stretching of turbulent fields in gas flows. The turbulent dynamo mechanism would be able to generate and preserve these large scale coherent magnetic fields with the appropriate spiral shape (Beck 2005b and references therein). Also, the presence of magnetic arms between gas arms in most of the spiral galaxies, provides an extra support for the global dynamo hypothesis; for if B were produced essentially by stellar winds, jets and SN explosions then one would expect much more intense fields within the arms where these phenomena are more frequent.

A survey of 74 spiral galaxies (see Beck 2005b and references therein) has revealed average magnetic fields $<B> = 9$ µG. In starburst (SB) galaxies, i.e., galaxies with high rates of star formation, the average field intensity is even larger, $B \geq 30\text{-}50\mu G$ and in the nuclear regions of these galaxies (e.g., NGC1067), magnetic fields $B \geq 100\mu G$ have been detected. This suggests a correlation between the magnetic fields and star formation in these galaxies. The magnetic fields may play an important role in the deflagration of SBs and, since these galaxies are believed to be the progenitors of spheroid galaxies, hence in the formation of spheroid galaxies (Totani 2001). According to Totani, this correlation could be due to magnetic braking, i.e., in order to the protogalactic cloud to collapse a significant amount of angular momentum must be transferred outwards. Then he suggests that SBs (and hence massive galaxy formation) would take place only where $B$ is strong enough to provide magnetic braking and allow the protogalaxy collapse at a critical value that happens to be of the order of few micro Gauss.

The magnetic energy density in the inner disk of some galaxies has been found to be larger than the thermal energy density, comparable to that of turbulent gas motions (another result that is consistent with dynamo action), and dominant in the outer disk (e.g., NGC6946). These evidences led Battaner & Florido (2000) to investigate the dynamics of these galaxies and they have found that magnetic field forces at the outside regions cannot be neglected in these cases and are a competitive candidate to explain the observed *flatness* of rotation curves without the necessity of invoking the presence of dark matter.

In some galaxies with high star formation rate it has been detected the presence of magnetic filaments and loops that are coupled with charged dust and rise above the galactic disk into the halo up to heights of ~5 kpc or more (e.g., NGC891; Dettmar 2005). They resemble the magnetic structures observed at the solar surface and suggest the presence of similar magnetic activity. Supernova shock fronts can break through the galactic disk and inject hot gas into the galactic halo, perhaps driving turbulence and the magnetic fields into it (e.g., Melioli & de Gouveia Dal Pino 2004; Melioli, de Gouveia Dal Pino, D'Ercole, Brighenti, Raga, 2005; Kim et al. 2006).

In several galaxies, the halo gas is observed to rotate slower than the gas in the disk (NGC5775). This velocity gradient could contribute to the excitation of a global dynamo in disk galaxies in a similar way to the differential rotation in the solar dynamo (Dettmar 2005).

## MAGNETIC FIELDS IN CLUSTERS OF GALAXIES AND THE IGM

Looking into the large scale structure of the Universe, we see that galaxies tend to aggregate into clusters. The Milky Way, for instance, belongs to the Virgo Cluster. A cluster has a typical diameter of ~100 Mpc (3 million light-years) and can be regular or irregular. Faraday rotation measurements of polarized Synchrotron radiation from embedded or background radio galaxies indicate magnetic fields in the Abel clusters $B\sim2\mu G$ with coherent length scales $L\sim10$ kpc (e.g., Grasso & Rubinstein 2001). In the central regions of clusters with embedded radio galaxies the magnetic fields are even larger: $B\sim5\mu G$ in irregular clusters, and $B\sim10\text{-}30\mu G$ in regular ones (Govoni et al. 2005), which are of the order of the magnetic field strengths of the galaxies. In some cluster cores, there seems to be some evidence that the magnetic pressure dominates the thermal pressure. Observations also reveal that the magnetic fields in the intra-cluster medium are predominantly turbulent with a power law spectrum which is approximately Kolmogorov at least at the small scale structures ($\leq 1$ kpc) (e.g., Hydra Cluster; Ensslin, Vogt & Pfrommer 2005; Govoni et al. 2005).

These magnetic fields may have been powered by: galactic winds from starburst (SBs) galaxies, jets from radio galaxies, cluster mergers, or dynamo action (Ensslin, Vogt & Pfrommer 2005 and references therein). Colgate & Li, have proposed that a dynamo operating in the accretion disk around the massive black hole (BHs) in the nucleus of radio galaxies could produce magnetized jets and inject them into the intra-cluster medium. Numerical Simulations of Kato et al. (2004) of magnetic field generation from accretion disks have shown that this is possible. However, the dynamics of the process leading to the formation of massive BHs in the nuclei of the active galaxies is still unclear and a pre-existing magnetic field may be required to carry away the huge angular momentum of the accreting matter. Besides, radio sources can generate only $B\approx10^{-7}$G (Kronberg 2005), so that further amplification would be required and could be provided by, e.g., shocks produced by galaxy interactions. Likewise, SB galaxies could also produce only $B\approx5 \times 10^{-9}$ G that should be amplified by other mechanisms (Kronberg et al. 1999). Alternatively, Ensslin, Vogt & Pfrommer (2005) have tried to explain the fields in core of clusters in terms of turbulent dynamo action, where instead of differential rotation the flow is in a turbulent medium. In this model, dynamo action occurs in all scales and micro-dynamos will compose to develop large scale magnetic fields. Magnetic diffusivity constrains the magnetic field scale to a fraction of the turbulent scale $L_B \propto L_T (R_M)^{-1/2}$, where $R_M$ is the magnetic Reynolds number. As $B$ grows, it reacts and tends to untangle thus decreasing $R_M$ and increasing $L_B$ and generating a more organized $B$.

Magnetic fields probably pervade the entire Universe. The diffuse intergalactic medium (IGM) is very rarefied but Faraday rotation of polarized emission from distant quasars (up to z=2.5) indicate that $B_{IGM} \leq 10^{-9}$ G, assuming a coherent length of the order of the maximum one measured in clusters $L \approx 1$ Mpc (e.g., Grasso & Rubinstein 2001). Where and how do these magnetic fields originate? Are they primordial?

## MAGNETIC FIELDS IN THE EARLY UNIVERSE

The history of the Universe started about 15 billion years ago with the Big-Bang singularity and since then it has expanded and cooled down, and suffered several phase transitions. When its temperature was about $10^9$ K and its size ~100 pc, the primordial nucleosynthesis took place with the formation of the first H and He nuclei. When the Universe was about 0.5 Myr old, its temperature decreased to 10,000 K and the electrons and protons recombined to form neutral H atoms and the radiation decoupled from matter. This relic radiation from the Big-Bang has been cooling since then and presently permeates the whole cosmos with a microwave background radiation (CMBR) of temperature $T$=3 K. The recombination was followed by a *dark age* where possibly the *first objects* in the Universe were formed and these caused a *reionization* of it when it was about 0.1 Gyr old. This epoch was probably particularly favourable for creation of magnetic fields (see below). The reionization was followed by the formation of the galaxies and the larger structures in the Universe (at ~1 Gyr).

Both the CMBR and the Big-Bang nucleosynthesis (BBN) impose constraints on the strength of primordial fields (e.g., Shaposhnikov 2005; Grasso & Rubinstein 2001). In the case of the BBN, the presence of a strong magnetic field could change the expansion rate of the Universe and thus the abundance of the primordial elements. The observed cosmic helium abundance implies an upper limit for the magnetic field at the time of the BBN, $B <10^{11}$ G which expanded to the present Universe (under the hypothesis of magnetic flux conservation and performing a proper statistical average of small flux elements; e.g. Grasso & Rubinstein 2001) would imply a present field permeating the IGM, $B_o <10^{-10}$ G (for coherent lengths of 1 Mpc). In the case of the CMBR, the existence of a strong magnetic field at the recombination era would have affected it through two effects: (i) breaking the spatial isotropy; and (ii) producing MHD alterations on the temperature and polarization fluctuations of the CMBR. These fluctuations have been imprinted on the CMBR by primordial density fluctuations (probably produced in the Universe right after the B-B, during the inflation phase, and later on allowed the formation of the large structures in the Universe). The presence of a strong magnetic pressure could, for example, reduce the acoustic peaks of these fluctuations by opposing the infall of photons and baryons into their gravitational well. The lack of these effects on the CMBR implies an upper limit on the background magnetic field at the present Universe $B_o< 10^{-8}$ –$10^{-9}$ G. Thus, these upper limits derived from the CMBR and BBN constrain the present cosmic magnetic field to values that are consistent with the inferred values for the IGM today ($10^{-9}$ G.). Also, it is possible to demonstrate that the diffusion time scale of these primordial fields is much larger than the age of the Universe and the corresponding diffusion length of these magnetic fields is $<10^9$ cm, so that any magnetic fields produced in the early Universe with present length scales larger than $10^9$ cm would have survived and probably left no significant imprints on the BBN or CMBR !

Is the inferred IGM magnetic field above primordial? As remarked in the previous sections, the origin of the galactic and cluster magnetic fields is probably due to the amplification of *seed* fields via galactic dynamo or more recent processes. But where and when were these first fields produced? The existing models for primordial magnetic fields are still rather speculative. There are those that propose more recent *astrophysical* origins for the magnetic fields and their seeds. Among these, one can mention the Biermann battery in intergalactic shocks, the Harrison effect, stellar magnetic fields, supernova explosions, galactic outflows into the intergalactic medium, and jets from active galaxies; and there are those models called cosmological ones, that invoke sometimes peculiar processes in the very early Universe (prior to the recombination era) to create seed magnetic fields. I will mention below only some examples of these models (and recommend, e.g., Grasso & Rubinstein 2001; Shaposhnikov 2005; Grasso 2006; Rees 2006 for those interested in a more complete review of the subject).

Among the cosmological models, Turner & Widrow (1988) investigated the possibility of magnetic field production during the inflation phase and found that if conformal invariance of the electromagnetic field was broken then a magnetic field would be created from super-adiabatic amplification of pre-existing quantum fluctuations and would have a value at the present universe $B_o(1Mpc) \approx 10^{-62}$ G, which is too small even to seed a galactic dynamo. An alternative mechanism could have operated during the QCD phase transition when the Universe was at a temperature $T_{QCD}$= 1.5 $10^{12}$ K. At this temperature, quarks start combining to form hadrons. As $T$ cools below $T_{QCD}$, bubbles with hadrons grow as deflagration burning fronts and become supersonic; the shock then re-heats the plasma stopping momentarily the bubble growth. The transition ends when the expansion wins and the quark pockets are completely hadronized. Electric fields develop behind the shock fronts, ahead of the expanding bubbles, and currents and magnetic fields $B \approx Ev$ are generated in small scales. The volume average of these fields is too small, $B_{QCD} = 10^8$ G, for producing significant relic fields in the present Universe, however, under very special conditions, such as an efficient amplification by hydromagnetic instabilities, dynamo operation during the QCD, and inverse cascade, Sigl et al. (1997) have found that this field could have amplified to $B_{QCD} = 10^{17}$ G and would imply a today's field $B_o(100 \text{ kpc}) \approx 10^{-9}$ G which is of the order of the predicted IGM field.

Among more recent, post-recombination candidates, the Biermann battery appears as a promising mechanism for seed field generation. It is well known that a third term should appear in the magnetic induction equation (eq. 1) whenever the flow has pressure and density gradients which are non parallel ($\nabla p \times \nabla n \neq 0$). During the re-ionization era, this mechanism may have created pre-galactic seed fields $B_o \approx 10^{-21}$ G that were then exponentially amplified by dynamo action within the proto-galactic clouds that originated the galaxies later on. Another possibility is the Harrison mechanism, originally proposed by Harrison (1970) to operate in proto-galactic structures, it could also have operated in pre-recombination plasmas where vortices would be present. The essential idea of this mechanism is that in a primordial flow with vortical structures, protons will rotate (through collisions) with the flow, but the electrons will be Thomson scattered by the photons of the CMBR. This will cause charge separation and generate electric currents and thus magnetic fields. It can be shown that the operation of this mechanism within vortices in the pre-recombination plasma would be able to create seeds with $10^{-18}$ G with correlation lengths of 1 Mpc in the present IGM. These could be amplified by dynamo action within the galaxies and provide the observed galactic magnetic fields (~few $\mu$G). A recent interesting suggestion, which is alternative to the dynamo mechanism, is that seed fields could be amplified by supernova-driven turbulence. Kim et al. (2006) argue that the observation of magnetic fields of $\mu$G levels already in very distant galaxies which were formed very long ago (at high redshifts, $z \approx 2$) shorten the time available for dynamo action. In order to amplify a seed field by the alpha-omega dynamo mechanism, which requires ~few Gyr growth time, the strength of the seed field should be stronger than $10^{-11}$ G. On the other hand, the growth time of magnetic fields amplified by SN-driven turbulence is only ~ 10 Myr, for 100 pc scales. This suggests a more efficient mechanism for kpc-scale magnetic field amplification, which should, however, still be confirmed by numerical experiments with hugh computational domains.

## FUTURE NEEDS AND PERSPECTIVES

Advances are needed not only in the theory and magneto-hydrodynamical multidimensional numerical studies, but perhaps even more urgent is the requirement for polarization observations at higher angular resolution in order to map the full wealth of magnetic structures in galaxies, the ISM and the IGM, that will in turn provide inputs for dynamo and MHD models. In our Galaxy, for instance, we need polarization maps of selected regions at high frequencies and a much larger sample of rotation measure data from pulsars (Beck 2005b).

There is presently a project for the construction of a next-generation radio telescope array that will have a total effective collecting area of 1 km$^2$ (with sensitivity at the frequencies 100 MHz to 25 GHz), the so called Square Kilometer Array (SKA). It will be composed of 150 stations of 100 m diameter – accounting for half the SKA area – that will be distributed across continental distances (~3000 km), and the remaining area (called core) will be concentrated within a central region of 5 km diameter. It will be built under an international collaboration involving several European countries, Australia, South Africa, China, USA, Argentina and Brazil. Cosmic magnetism is one of the key science projects of the SKA. It will be able to detect more than 10,000 pulsars in our Galaxy which will allow to map the large scale spiral structure of its magnetic field. Besides, the SKA will be able to map nearby galaxies with at least 10 times better angular resolution compared to present-day radio telescopes, or 10 times more distant galaxies with similar spatial resolution as today. Magnetic field structures will illuminate the dynamical interplay of cosmic forces. The SKA's sensitivity will allow to detect synchrotron emission from the most distant galaxies in the earliest stage of evolution and to search for the earliest magnetic fields and their origin (Beck 2005b).

## ACKNOWLEDGMENTS

The author would like to thank the invitation of the organizers of the XI-LAWP for delivering this lecture at the Conference and also the kind hospitality of the staff of the Instituto de Fisica Nuclear of UNAM, specially of Alex Raga. The author also acknowledges partial support of a grant of the Brazilian Research Council (CNPq) and from UNAM.

# REFERENCES

1. F. Bacciotti, T. P. Ray, R. Mundt, J. Eislöffel, and J. Solf, , ApJ, 2002, 576, 222.
2. E. Battaner, and E. Florido, Fundamentals of Cosmic Fields, 2000, 21, 1-154.
3. R. Beck, in *Magnetic Fields in the Universe: from Laboratory and Stars to Primordial Structures*, edited by E. M. de Gouveia Dal Pino, G. Lugones and A. Lazarian, AIP Conference Proceedings 784, American Institute of Physics, Melville, NY, 2005a, pp. 453-355.
4. R. Beck, in *Magnetic Fields in the Universe: from Laboratory and Stars to Primordial Structures*, edited by E. M. de Gouveia Dal Pino, G. Lugones and A. Lazarian, AIP Conference Proceedings 784, American Institute of Physics, Melville, NY, 2005b, pp. 343-353.
5. Blandford, R.D. & Payne, D.G. MNRAS, 1982, 199.
6. H. Carmichael, in *Proc. of AAS-NASA Symp. on the Physics of Solar Flares*, Edited by W. N. Hess, NASA-SP 50, p. 451.
7. R. M. Crutcher, in *Magnetic Fields in the Universe: from Laboratory and Stars to Primordial Structures*, edited by E. M. de Gouveia Dal Pino, G. Lugones and A. Lazarian, AIP Conference Proceedings 784, American Institute of Physics, Melville, NY, 2005, pp. 129-139.
8. E. M. de Gouveia Dal Pino, in *Magnetic Fields in the Universe: from Laboratory and Stars to Primordial Structures*, edited by E. M. de Gouveia Dal Pino, G. Lugones and A. Lazarian, AIP Conference Proceedings 784, American Institute of Physics, Melville, NY, 2005a, pp. 183-194.
9. E. M. de Gouveia Dal Pino, Advances in Space Research, 2005b, 35, 908-924
10. E. M. de Gouveia Dal Pino, 2006a, Braz. J. Phys. (in press)
11. E. M. de Gouveia Dal Pino, in *Cosmic Magnetic Fields*, edited by R. Beck et al., 2006b, Astron. Nachr, 2006 (in press).
12. E. M. de Gouveia Dal Pino and A. Lazarian, A&A, 2005, 441, 845-853.
13. R.-J. Dettmar, in *Magnetic Fields in the Universe: from Laboratory and Stars to Primordial Structures*, edited by E. M. de Gouveia Dal Pino, G. Lugones and A. Lazarian, AIP Conference Proceedings 784, American Institute of Physics, Melville, NY, 2005a, pp. 354-361.
14. T. A. Ensslin, C. Vogt, and C. Pfrommer, in *The Magnetized Plasma in Galaxy Evolution,* edited by K. T. Chyzy et al., Jagiellonian University, Krakow, 2005, 231-238.
15. F. Govoni et al., in *Cosmic Magnetic Fields*, edited by R. Beck et al., 2006, Astron. Nachr, 2006 (in press).
16. D. Grasso, and H.R. Rubinstein, Physics Reports, 2001, 348, 163-266.
17. D. Grasso, in *Cosmic Magnetic Fields*, edited by R. Beck et al., 2006, Astron. Nachr, 2006 (in press).
18. G. A. Guerrero and E. M. de Gouveia Dal Pino, 2006 (in prep.).
19. E.R. Harrison, Mon. Not. R. Astron. Soc., 1970, 147, 279.
20. Y. Kato, S., Mineshige, and K. Shibata, ApJ, 2004, 605, 307.
21. J. Kim et al., in *Cosmic Magnetic Fields*, edited by R. Beck et al., 2006, Astron. Nachr, 2006 (in press).
22. A. Königl, A., and R. Pudritz, in *Protostars and Planets IV*, (Book - Tucson: University of Arizona Press); edited by Mannings, V., Boss, A.P., Russell, S. S., 2000, p. 759.
23. P. P. Kronberg, in in The Magnetized Plasma in Galaxy Evolution, edited by K. T. Chyzy et al., Jagiellonian University, Krakow, 2005, 239-246.
24. P. P. Kronberg, H. Lesch and U. Hopp, ApJ, 1999, 511, 56.
25. Zhi-Yun, Li, and F. H. Shu, ApJ, 1996, 468, 261.
26. A. M. Magalhães et al., in *Magnetic Fields in the Universe: from Laboratory and Stars to Primordial Structures*, edited by E. M. de Gouveia Dal Pino, G. Lugones and A. Lazarian, AIP Conference Proceedings 784, American Institute of Physics, Melville, NY, 2005, pp. 715-720.
27. C. Melioli, E.M. de Gouveia Dal Pino, and A. Raga, A&A, A&A, 2005, 443, 495-508.
28. C. Melioli, A. D'Ercoli, E.M. de Gouveia Dal Pino, F. Brighentti, and A. Raga, 2006 (in prep.)
29. M. Rees, in *Cosmic Magnetic Fields*, edited by R. Beck et al., 2006, Astron. Nachr, 2006 (in press).
30. M. Shaposhnikov, in *Magnetic Fields in the Universe: from Laboratory and Stars to Primordial Structures*, edited by E. M. de Gouveia Dal Pino, G. Lugones and A. Lazarian, AIP Conference Proceedings 784, American Institute of Physics, Melville, NY, 2005, pp. 423-433.
31. K. Shibata, in *Magnetic Fields in the Universe: from Laboratory and Stars to Primordial Structures*, edited by E. M. de Gouveia Dal Pino, G. Lugones and A. Lazarian, AIP Conference Proceedings 784, American Institute of Physics, Melville, NY, 2005, pp. 153-163.
32. G. Sigl, A. Olinto, K. Jedamzik, Phys. Rev. D 55 (1997) 4582.
33. A. V. R. Silva, in *Magnetic Fields in the Universe: from Laboratory and Stars to Primordial Structures*, edited by E. M. de Gouveia Dal Pino, G. Lugones and A. Lazarian, AIP Conference Proceedings 784, American Institute of Physics, Melville, NY, 2005, pp. 90-102.
34. K. Subramanian, MNRAS, 1998, 294, 718.
35. T. Totani, [astro-ph/9904042].
36. M. S. Turner and L. M. Widrow, Phys. Rev. D, 1988, 37, 2743.